\documentclass[a4paper]{aa}
\usepackage[varg]{txfonts}
\pdfoutput=1

\usepackage{hyperref}
\hypersetup{colorlinks=true,linkcolor=blue,citecolor=blue,filecolor=blue,urlcolor=blue}

\usepackage{graphicx}
\graphicspath{{images/}}

\usepackage{bm}

\usepackage{booktabs,dcolumn}
\newcolumntype{d}[1]{D{.}{.}{#1}}
\newcommand{\mcc}[1]{\multicolumn{1}{c}{#1}}
\newcommand{\mccc}[1]{\multicolumn{2}{c}{#1}}

\newcommand{\msun}{M_\odot}
\newcommand{\der}{\mathrm{d}}
\newcommand{\tcr}{t_\mathrm{cr}}
\newcommand{\trh}{t_\mathrm{rh}}
\newcommand{\tend}{t_\mathrm{end}}
\newcommand{\tbh}{t_\mathrm{BH}}
\newcommand{\tot}{\mathrm{tot}}
\newcommand{\rh}{r_\mathrm{h}}
\newcommand{\rp}{r_\mathrm{P}}
\newcommand{\rv}{r_\mathrm{v}}
\newcommand{\rt}{r_\mathrm{t}}

\newcommand{\pc}{\mathrm{pc}}
\newcommand{\kpc}{\mathrm{kpc}}
\newcommand{\Myr}{\mathrm{Myr}}
\newcommand{\kms}{\mathrm{km\,s^{-1}}}

\newcommand{\rtf}{\eta_\mathrm{BH}}
\newcommand{\disp}{\sigma_\mathrm{BH}}
\newcommand{\vkick}{v_\mathrm{kick}}
\newcommand{\vki}{v_\mathrm{kick,i}}

\newcommand{\vesc}{v_\mathrm{esc}}
\newcommand{\menv}{m_\mathrm{env}}
\newcommand{\mbh}{m_\mathrm{BH}}

\renewcommand{\vec}[1]{\overrightarrow{\bm{#1}}}

\begin{document}

\title{The black hole retention fraction in star clusters}
\subtitle{}

\setcounter{footnote}{4}

\author{Václav Pavlík\inst{\ref{auuk},\ref{obs},}\thanks{\email{pavlik@sirrah.troja.mff.cuni.cz}}
\and Tereza Jeřábková\inst{\ref{auuk},\ref{bonn},\ref{eso},}\thanks{\email{tjerabko@eso.org}}
\and Pavel Kroupa\inst{\ref{auuk},\ref{bonn}}
\and Holger Baumgardt\inst{\ref{uq}}}

\institute{Astronomical Institute of Charles University, Prague, Czech Republic\label{auuk}
\and Observatory and Planetarium of Prague, Prague, Czech Republic\label{obs}
\and Helmholtz-Institut für Strahlen- und Kernphysik (HISKP), Universität Bonn, Bonn, Germany\label{bonn}
\and European Southern Observatory, Garching bei München, Germany\label{eso}
\and School of Mathematics and Physics, University of Queensland, St. Lucia, QLD 4072, Australia\label{uq}}

\titlerunning{The black hole retention fraction in star clusters}
\authorrunning{Pavlík et al.}

\date{February 28, 2018 / May 28, 2018}

\abstract
{Recent research has been constraining the retention fraction of black holes (BHs) in globular clusters by comparing the degree of mass segregation with $N$-body simulations. They are consistent with an upper limit of the retention fraction being 50\,\% or less.}
{In this work, we focus on direct simulations of the dynamics of BHs in star clusters. We aim to constrain the effective distribution of natal kicks that BHs receive during supernova (SN) explosions and to estimate the BH retention fraction.}
{We used the collisional $N$-body code \texttt{nbody6} to measure the retention fraction of BHs for a given set of parameters, which are: the initial mass of a star cluster, the initial half-mass radius, and $\disp$, which sets the effective Maxwellian BH velocity kick distribution. We compare these direct $N$-body models with our analytic estimates and newest observational constraints.}
{The numerical simulations show that for the one-dimensional (1D) velocity kick dispersion $\disp < 50\,\kms$, clusters with radii of 2\,pc and that are initially more massive than $5 \times 10^3\,\msun$ retain more than 20\,\% of BHs within their half-mass radii. Our simple analytic model yields a number of retained BHs that is in good agreement with the $N$-body models. Furthermore, the analytic estimates show that ultra-compact dwarf galaxies (UCDs) should have retained more than 80\,\% of their BHs for $\disp \leq 190\,\kms$. Although our models do not contain primordial binaries, in the most compact clusters with $10^3$ stars, we have found evidence of delayed SN explosions producing a surplus of BHs compared to the IMF due to dynamically formed binary stars. These cases do not occur in the more populous or expanded clusters.}
{}


\maketitle

\section{Introduction}

Black holes (BHs) and their retention fraction in star clusters play an important role in the evolution of the clusters, and are relevant for other astrophysical fields including stellar evolution, the formation of intermediate-mass/super-massive BHs and predictions of gravitational wave events \citep[e.g.][]{Belczynski2002,Favata2004, Mackey2007,Banerjee2010,Mapelli2011,Mapelli2013,Morscher2013,Ziosi2014,Banerjee2017,Repetto2017,Banerjee2018}.  
The retained number of BHs also constrains the maximum mass of potentially formed intermediate-mass/super-massive BHs in massive stellar systems such as in globular clusters (GCs) and ultra-compact dwarf  galaxies (UCDs) \citep{jerabkova}.

Black holes can be observed and quantified either directly via gravitational waves or indirectly via (i) mass accretion (which requires the presence of gas and/or stellar companions), (ii) gravitational lensing, or (iii) mass segregation in star clusters \citep{baumgardt_sollima,mass_segr}.
Method (iii) has been successfully used to establish an upper limit on the BH retention fraction in GCs to be 50\,\% or less \citep{peuten,baumgardt_sollima}. However, the constraints on the initial retention fraction of BHs remain weak. By the initial retention, which is what this study is concerned with, we mean the fraction of BHs that are retained in the star cluster by the time of the last core collapsed supernova (SN) leaving a BH. This corresponds to a time-scale of about 12\,Myr after the formation of the cluster.

An asymmetric SN explosion gives the newly formed remnant an initial momentum, $\vec{p}_\mathrm{BH}$, to compensate for the excess momentum of the stellar envelope going in the opposite direction, $\vec{p}_\mathrm{env-}$, that is,
\begin{equation}
        \label{eq:momentum}
        \vec{p}_\mathrm{env-} = \vec{p}_\mathrm{env+} + \vec{p}_\mathrm{BH} \,,
\end{equation}
where $\vec{p}_\mathrm{env+}$ is the momentum vector of the envelope in the same direction as $\vec{p}_\mathrm{BH}$. The \emph{velocity kick}, $\vkick$, is then the initial remnant's velocity derived from its momentum and mass \citep{lyn94}.

Here we provide a systematic direct $N$-body survey of the initial retention fractions of BHs for different assumptions on the kick velocities for a variety of star cluster radii and masses, that is,\ in the range from $10^3$ to $10^5$ stars. We also compare these numerical results with our own analytic estimates on the retention fraction of BHs. Finally, we use this study to extrapolate to the initial retention fraction in larger systems, that is,\ GCs and UCDs.

\section{Numerical models}

We performed over 1\,500 $N$-body simulations of \emph{isolated} star clusters. Clusters with lower numbers of stars ($N < 25$k) were evolved with a single-central-processing-unit (CPU) \texttt{nbody6} integrator. Clusters with more stars were integrated with a parallel \texttt{nbody6.sse} version \citep{nbody6,nbody6_sse}\footnote{Both \texttt{nbody} versions used here are from June 13, 2017.}\!.
All the initial conditions were set up with a random seed by the integrator as follows.

The models presented contain 1k, 3k, 10k, 25k, 50k, and 100k stars; see Table~\ref{tab:models} for their detailed integration parameters.
We assumed the canonical initial mass function \citep[IMF,][]{kroupa01,kroupa13} with masses in the range from $0.08\,\msun$ to $100\,\msun$ and the distribution of positions of stars according to \citet{plummer} and \citet{plummer_method}. For each model, we used three values of the initial virial radius, $\rv = 0.5$, $1,$ and $2\,\pc$; the value of the initial Plummer radius is $\rp = \frac{3\pi}{16}\,\rv$ and the half-mass radius is $\rh \approx 1.305\,\rp$ \citep{kroupa08_lecture_notes}. Neither primordial binaries nor a gas component were included in the models.

For the stellar evolution, we invoked the algorithm by \citet{sse} for single stars and \citet{bse} for binary stars.
For that we assumed a metallicity $\left[ \mathrm{Fe}/\mathrm{H} \right] = -1.30$ which corresponds to the average metallicity of GCs in the Galaxy \citep{baumgardt_sollima}.
We have also performed one additional calculation of a vastly expanded star cluster ($\rv = 20.0\,\pc$) with the same metallicity and IMF and no primordial binaries in order to estimate the time scale needed for a single star to evolve into a BH. In an extended cluster like this, dynamical effects and binary evolution may be neglected. The last BH appeared at $\approx 11.7\,\Myr$ from the beginning of the integration. Therefore, all our simulations were closely followed over that time.

Finally, in each set-up, we assumed a different value for the one-dimensional (1D) dispersion, $\disp$,  the Maxwellian distribution from which we drew the initial BH kick velocities, $\vki$. It has been suggested, for example by \cite{bh_ns_kicks} or \citet{repetto2012}, that BHs may receive kicks as high as neutron stars (NSs), therefore the highest value of $\disp$ comes from the best fit of the velocity distribution of the observed NSs \citep{kicks_ns}, that is,\ $\disp = 190\,\kms$. We set the lowest value of $\disp$ to $3\,\kms$ and a moderate value to $50\,\kms$. In the case of our 1\,pc models, we also included $\disp = 10$ and $25\,\kms$; see Table~\ref{tab:models}. The kick velocity is composed of three random deviates chosen from a Gaussian distribution with a 1D dispersion $\disp$ \citep[e.g.][]{kroupa08_lecture_notes}.

\begin{table}
  \centering
  \caption{Parameters of the star-cluster models used in this paper. For each number of stars (with an approximate total mass), there are the initial virial radius, half-mass relaxation time, crossing time, and the approximate time up to which we followed the integration. The estimated time of formation of the last BH in our models is $\approx 11.7\,\Myr$.}
  \begin{tabular}{cd{2}d{2}d{3}d{2}}
    \toprule
    $N$ & \mcc{$\rv\ [\pc]$} & \mcc{$\trh\ [\Myr]$} & \mcc{$\tcr\ [\Myr]$} & \mcc{$\tend\ [\Myr]$} \\
    ($M_\tot [\msun]$) &&&&\\
    \midrule
    1k                                  & 0.5 & 6.1 & 0.61  & 65  \\
    ($5 \times 10^2$)   & 1.0 & 19  & 1.9   & 201 \\
                                        & 2.0 & 49  & 4.9   & 523 \\
    &&&&\\
    3k                                  & 0.5 & 9.3 & 0.36  & 38  \\
    ($1.6 \times 10^3$) & 1.0 & 27  & 1.0   & 110 \\
                                        & 2.0 & 75  & 3.0   & 313 \\
    &&&&\\
    10k                                 & 0.5 & 15  & 0.20  & 35  \\
    ($5.5 \times 10^3$) & 1.0 & 42  & 0.56  & 99  \\
                                        & 2.0 & 118 & 1.6   & 280 \\
    &&&&\\
    25k                                 & 0.5 & 21  & 0.13  & 23  \\
    ($1.4 \times 10^4$) & 1.0 & 60  & 0.36  & 64  \\
                                        & 2.0 & 168 & 1.0   & 179 \\
    &&&&\\
    50k                                 & 0.5 & 28  & 0.088 & 16  \\
    ($2.8 \times 10^4$) & 1.0 & 79  & 0.25  & 45  \\
                        & 2.0 & 223 & 0.71  & 125 \\
    &&&&\\
    100k                                & 0.5 & 37  & 0.062 & 17  \\
    ($5.6 \times 10^4$) & 1.0 & 107 & 0.18  & 43  \\
                                        & 2.0 & 301 & 0.50  & 119  \\
    \bottomrule
  \end{tabular}
  \label{tab:models}
\end{table}

In this work, we do not treat a detailed kick velocity mechanism (we simply assume a Maxwellian distribution). There are currently different models of how the kicks are produced during the SN explosions:
After a SN explosion, the initial velocity kick, $\vki$, may be reduced by the mass that falls back onto the remnant. One possibility is that the kick velocity scales with the ratio between the mass of the envelope, $\menv$, and the mass of the star before the SN explosion as
\begin{equation}
        \label{eq:fallback_nb6}
        \vkick = \vki \frac{\menv}{\mbh + \menv} \,,
\end{equation}
where $\mbh$ is the remnant's mass \citep[][]{nbody6}. However, according to a recent study by \cite{belczynski} the remnant's velocity should be scaled by a fraction of the envelope mass, $\Delta\menv$, that is,
\begin{equation}
        \label{eq:fallback}
        \vkick = \vki \left( 1 - \frac{\Delta\menv}{\menv} \right) \,.
\end{equation}
\citet{fryer} use the same prescription for the final magnitude of the velocity kick and they also argue that the fall-back mass is proportional to the mass of the envelope, although they determined the mass differently. In either case, the recoil is given by the momentum conservation due to an asymmetric spatial distribution of the mass of the envelope; see eq.~\eqref{eq:momentum}, and the final velocity is scaled by the mass that falls back ($\Delta\menv$). Therefore, the total mass of the star \emph{before} a SN, as in eq.~\eqref{eq:fallback_nb6}, may not be a valid scaling parameter for the fall-back since it cannot help us establish by how much the fall-back fraction of the envelope actually slows the remnant.
Because in our case the masses of the BHs are similar, the effect of fall-back on the final kick velocity is comparable to evaluating different values of $\disp$. Therefore, in our models, there is no rescaling of the kick velocity, that is,\ $\vkick = \vki$.

In our calculations, if a star explodes as a SN while bound in a binary system, the remnant does not receive a kick. This is the case in less than 2\,\% of stars that lead to SNe, so this does not significantly affect the results. Because the kick velocity dispersion $\disp$ is an assumed quantity, it is to be interpreted as an effective kick velocity dispersion. This means that any physical process that changes the actual kick, $\vki$, to the velocity of the BH when it is free streaming in the cluster (e.g.\ after dissociating from a binary) is not considered
here.

\begin{figure}
  \centering
  \includegraphics{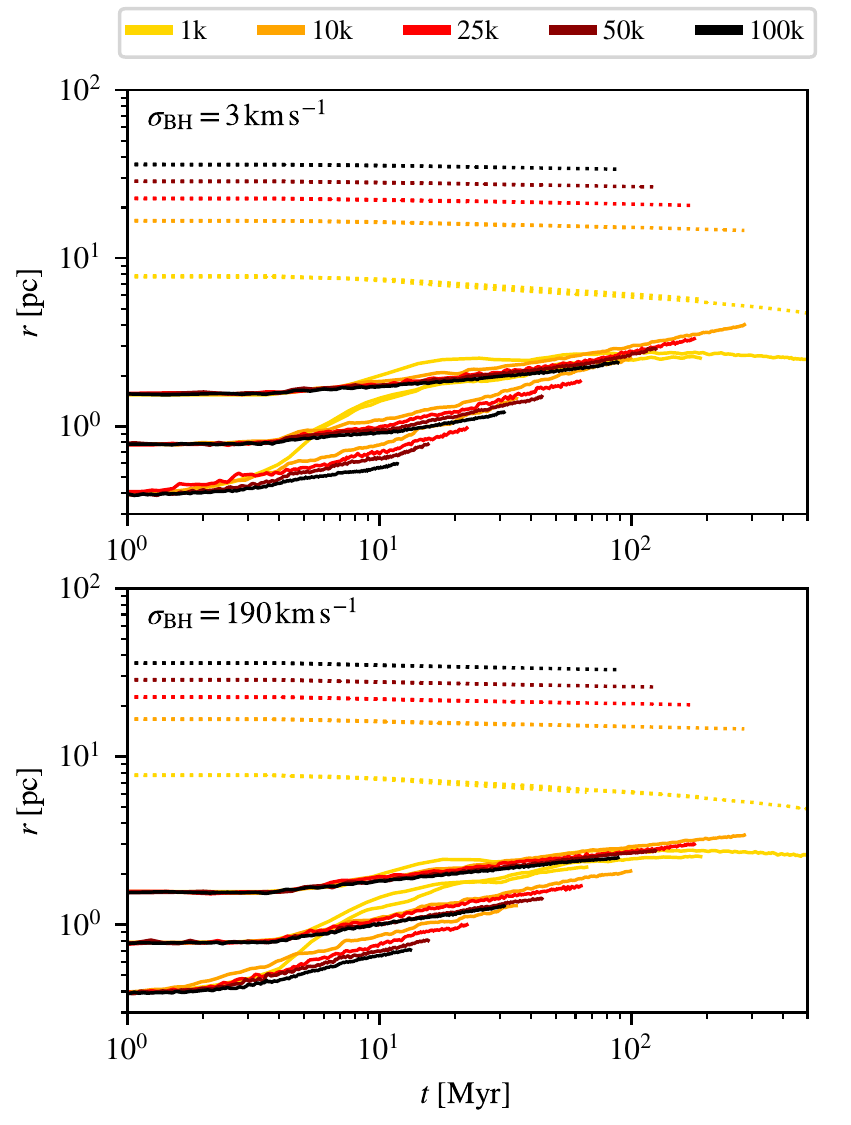}
  \caption{The evolution of the half-mass radii ($\rh \approx 0.769\,\rv$, solid lines) and tidal radii (dotted lines) in our models. We compare models with the lowest and highest $\disp$. Models with different N are colour-coded as given by the key at the top of the figure.}
  \label{fig:radii}
\end{figure}

We note that, according to \citet{bel_sse,2016ApJ...819..108B}, for example, BH masses may grow larger for lower metallicities than according to \citet{sse}. This, however, does not have any effect on our present calculations of the kick velocity because in our models, $\vkick$ does not depend on the BH mass. The only possible effect could be through the cluster's expansion as a result of larger masses of the lost BHs.
In the models presented, about 6\,\% of the mass of the cluster can be lost if all BHs escape. This value could increase by a maximum factor of two if we consider the approach of \cite{bel_sse}. Although this would be useful to study in the future, it should not have a significant effect on our results; see, for example, Fig.~\ref{fig:radii} where we compare the cluster's expansion by means of the half-mass and tidal radii. In the models with $\disp = 190\,\kms$, almost all BHs escaped, while in the models with $\disp = 3\,\kms$ only a small fraction did (see Sect.~\ref{sec:results}). Nevertheless, no significant difference between those two plots can be seen.

\section{Methods}

\subsection{Retention fraction}
\label{sec:rtf}

In each realisation of our models, we tracked the positions of all newly formed and existing BHs. The retention fraction, hereafter denoted as $\rtf$, corresponds to the fraction of BHs that are inside a certain radius, for example,\ the half-mass radius or the cut-off radius.

Both of these radii are derived with respect to the density centre, provided by \texttt{nbody6} \citep[according to][]{casertano_hut}. The cut-off radius, $\rt$, is taken to be the tidal radius \citep[e.g.][]{binney_tremaine}
\begin{equation}
  \label{eq:tidal}
  \rt = r_\mathrm{G} \left( \frac{M_\mathrm{C}}{3 M_\mathrm{G}} \right)^{\frac{1}{3}} \,,
\end{equation}
where $M_\mathrm{C}$ is the mass of the cluster within this radius, $r_\mathrm{G}$ is the distance of the cluster from the centre of the Galaxy (assumed to be $5\,\kpc$), and $M_\mathrm{G}$ is the mass of the Galaxy comprised in the radius $r_\mathrm{G}$. According to \citet{galaxy_mass} and \citet{gallaxy_mass_new}, we took $M_\mathrm{G} \approx 5\times10^{10}\,\msun$.
At each time step, the half-mass radius is calculated from the stars that are bound to the cluster, that is,\ up to $\rt$ from eq. \eqref{eq:tidal}.

Certain dynamical effects, for example,\ close encounters of two stars or single stars with binaries, are able to significantly reduce the fraction of BHs that are in the cluster on a timescale of hundreds of millions of years. Other effects, such as dynamical friction, can slow down escaping BHs sufficiently to maintain the BH population inside a cluster for longer.
In order to reduce the influence of these effects as much as possible, we evaluate the BH retention soon after the last BH has formed in each model (given the metallicity, this is at approximately 12\,Myr).
We are aware that the latest BHs with velocities drawn from a Maxwellian distribution with $\disp=3\,\kms$ may not be able to escape from the cluster by that time. On the other hand, their $\vkick$ would barely exceed $\vesc$, so our results should hold (see Sect.~\ref{sec:results} and Fig.~\ref{fig:dyn}).

\subsection{Analytic estimate}

To estimate the retention fraction analytically, we generated a set of star clusters with the canonical IMF in the same mass range as our $N$-body models. First, we assume that, at the time of the kick, the systems are \citet{plummer} models with the virial radius $\rv$ and that the kick velocities, $\vkick$, follow a Maxwellian distribution,
\begin{equation}
        \label{eq:pdf_maxwell}
    P(\vkick) = \sqrt{\frac{2}{\pi}} \, \frac{\vkick^2 \exp{\left( - \frac{\vkick^2}{2 \disp^2} \right)}}{\disp^3} \,,
\end{equation}
with a velocity dispersion $\disp$.
The values of $\disp$ used in these estimates are the same as we used in the initial conditions in the above $N$-body models.

The BH retention fraction is the fraction of BHs that do not escape from the cluster. Here, we assume that the stellar remnants are lost if their velocity after a SN kick is larger than the escape velocity at their current radius $r$ from the centre of the star cluster. This limit-velocity is defined as
\begin{equation}
  \label{eq:vesc}
  \vesc(r) = \left( \frac{2 G M_\mathrm{C}}{\sqrt{r^2 + \rp^2}} \right)^\frac{1}{2} \,,
\end{equation}
where $M_\mathrm{C}$ is the mass of the cluster, $G$ is the gravitational constant, and $\rp$ is the Plummer radius from the initial conditions of our models. The predicted ratio of retained BHs is then given by
\begin{equation}
  \label{eq:rtf_analytic}
  \rtf = \frac{\int_0^{\vesc}{P(v) \der v}}{\int_0^{\infty}{P(v) \der v}} \,.
\end{equation}

\section{Results}
\label{sec:results}

\subsection{Retention fraction}

If no stellar dynamics is taken into account, for a given effective kick velocity dispersion ($\disp$), the value of $\rtf$ should be stationary and depend only on the number of heavy stars in the model, that is,\ the initial cluster mass (given that we use the same IMF and stellar evolution in all models).
In Fig.~\ref{fig:rtf} and Table~\ref{tab:rtf}, we document the evolution of the mean retention fraction in our models. The results show that almost all BHs are kicked out of the clusters in our simulations with $\disp=190\,\kms$ and also that the retention fraction increases with a decreasing $\disp$. In Fig.~\ref{fig:radii}, we compare the temporal evolution of the half-mass and tidal (or cut-off) radii of our clusters with the lowest and highest $\disp$. The half-mass radii calculated from these two classes of models show not more than a marginal difference in the overall evolution. As the mass contained in the BHs is around 6\,\% of the total mass, it is expected that the whole cluster would not feel their absence.

\begin{figure}
  \centering
  \includegraphics{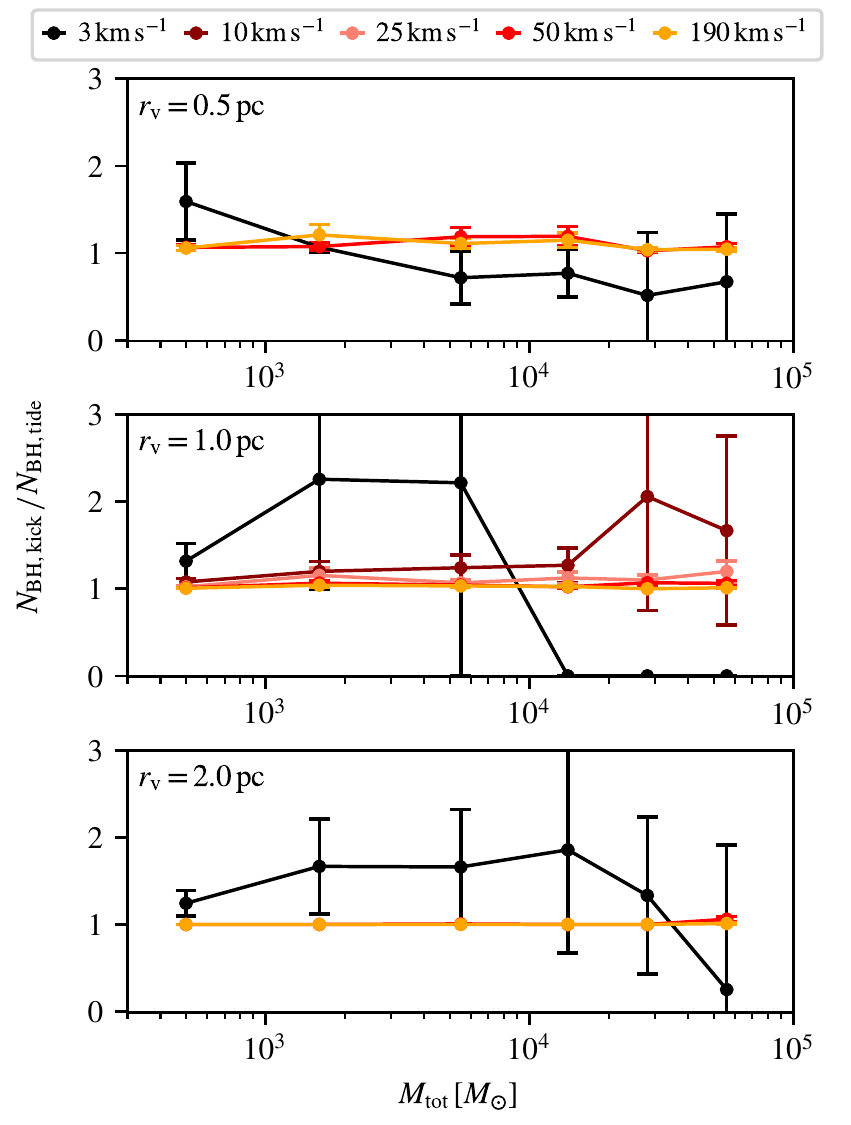}
  
  \caption{The average ratio between the number of BHs that received $\vkick > \vesc$ during a SN explosion ($N_\mathrm{BH,kick}$) and the number of BHs that are outside the tidal radius at 12\,Myr ($N_\mathrm{BH,tide}$). Each colour corresponds to a different kick velocity dispersion $\disp$ listed above the plots. The error bars correspond to the Poisson uncertainties; see eq. \eqref{eq:errors}.}
  \label{fig:dyn}
\end{figure}

We have also evaluated dynamical effects on the retention of BHs. In Fig.~\ref{fig:dyn} we plot the ratio
\begin{equation}
        \label{eq:dyn}
        Q \equiv \frac{N_\mathrm{BH,kick}}{N_\mathrm{BH,tide}} \,.
\end{equation}
Here, $N_\mathrm{BH,kick}$ is the number of BHs that received a kick with a higher velocity than the escape velocity at their current distance from the cluster centre from eq.~\eqref{eq:vesc}, that is,\ $\vkick > \vesc$. The denominator, $N_\mathrm{BH,tide}$, is the number of BHs that are outside the tidal radius of the cluster at 12\,Myr, that is,\ the approximate time necessary for BHs to form in our models. We also include the Poisson uncertainties defined as the square root of the value. Based on the propagation of uncertainty, the plotted error bars correspond to
\begin{equation}
        \label{eq:errors}
    \sigma_Q = \frac{N_\mathrm{BH,tide}\sqrt{N_\mathrm{BH,kick}} - N_\mathrm{BH,kick}\sqrt{N_\mathrm{BH,tide}}}{N_\mathrm{BH,tide}^2} \,.
\end{equation}
When confronted with the final $\rtf$, this ratio indicates one of three scenarios: all the BHs have been ejected because of a SN kick ($Q=1$), some of the kicked BHs were dynamically retained inside the cluster ($Q>1$), or how many BHs, not retained in the cluster, have been expelled dynamically through encounters with other stars or other BHs ($Q<1$).

We deduced that natal kicks are a dominant factor influencing the resulting initial retention fraction, $\rtf$, in our models with $\disp \geq 10\,\kms$. The only models where dynamical processes have a major influence on $\rtf$ are those with $\disp = 3\,\kms$ (black lines in Fig.~\ref{fig:dyn}), where $Q<1$. Nonetheless, even in those models, the retention fraction in the tidal radius is 82\,\% or higher (see Table~\ref{tab:rtf}). Ejecting a BH merely through dynamical processes is, therefore, rare on such a short timescale.

\begin{figure}
  \centering
  \includegraphics{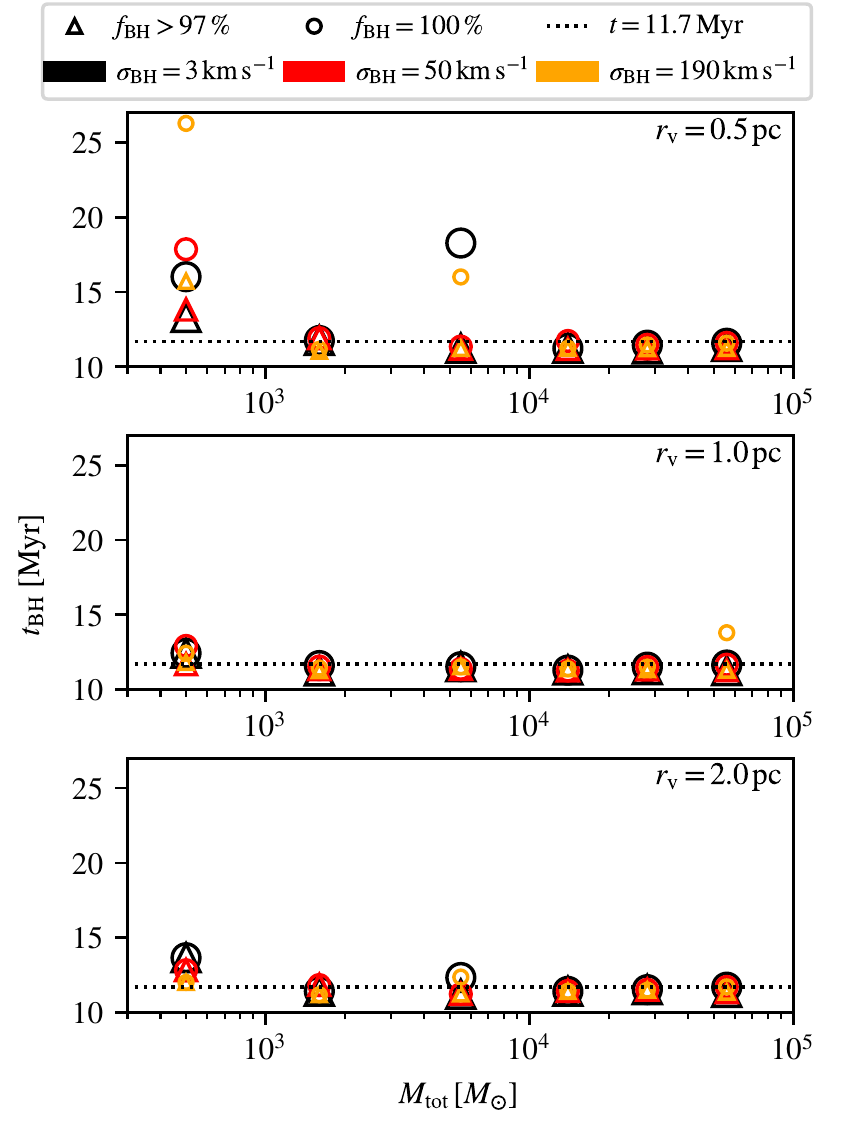}
  
  \caption{The time of BH formation in the presented models. Circles correspond to the time of the last BH formation (i.e.\ $f_\mathrm{BH} = 100\,\%$), triangles represent the time when a fraction $f_\mathrm{BH} > 97\,\%$ of BHs formed. For reference, the dotted line marks $11.7\,\Myr$, as deduced from the \citet{sse} algorithm.}
  \label{fig:last_bh}
\end{figure}

\begin{figure}
  \centering
  \includegraphics{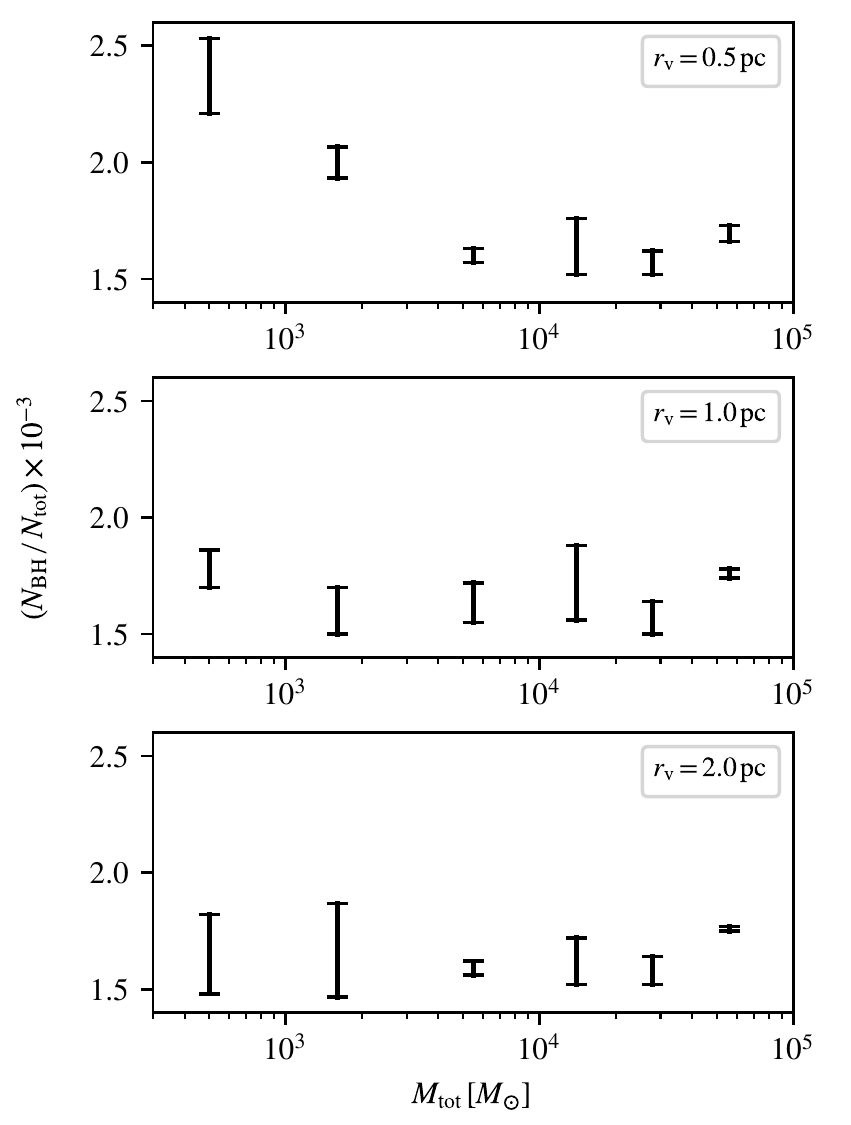}
  
  \caption{The ratio of BHs per total initial number of stars in our models. Each vertical bar goes from the lowest to the highest number of BHs found within the realisations.}
  \label{fig:mass}
\end{figure}

\begin{figure}
  \centering
  \includegraphics{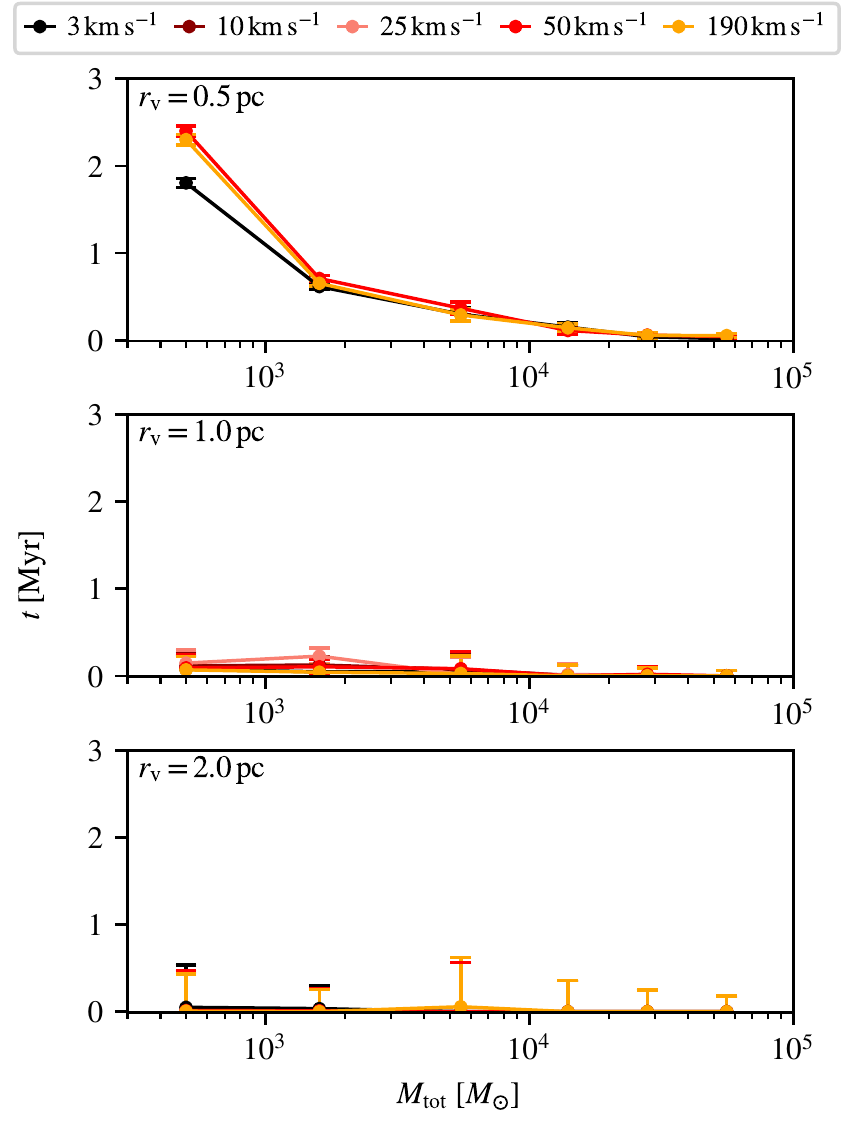}
  
  \caption{The mean time that a star, which ended up as a BH, spent in a binary (or multiple) system before becoming a BH. The vertical uncertainties represent one half of the output period.}
  \label{fig:bhbin}
\end{figure}

According to the single stellar evolution algorithm by \citet{sse} which is parametrised only by mass and metallicity of individual stars, the last SN explodes at $\approx 11.7\,\Myr$. For the small $N$ systems, we also see a systematic shift to a BH production beyond this time (in Fig.~\ref{fig:last_bh}). The reason is that the IMF sampled for 1\,000 stars gives only a couple of massive stars (sometimes only one). Due to dynamical processes, those few massive stars sink to the centre of the cluster and form a binary star that has a very small chance of being disrupted by the less-massive stars surrounding it. The binary stellar evolution algorithm \citep{bse} introduces additional parameters, for example,\ mass transfer, accretion, and collisions, which lead to a rejuvenation of the star, causing \emph{a delayed SN explosion}. This process is particularly pronounced, 
even by a factor of two (see the top panel of Fig.~\ref{fig:last_bh}),
for the $\rv=0.5\,\pc$ clusters in which the binaries that form are sufficiently compact for the rejuvenation to occur.
Individual cases, for example,\ the 100k model with $\rv=1.0\,\pc$ and $\disp=190\,\kms$ (orange circle in the middle panel of Fig.~\ref{fig:last_bh}), or the 10k model with $\rv=0.5\,\pc$ and $\disp=3$ and $190\,\kms$ (black and orange circles in the middle panel of Fig.~\ref{fig:last_bh}, respectively), also delayed the last BH formation to above 12\,Myr. We interpret those as being due to rejuvenated stars as well.

In the most compact models, that is,\ $\rv=0.5\,\pc$ with 1k (and also 3k) stars, we see an overproduction of BHs compared to other models (even with the same number of stars) --- compare the top panel of Fig.~\ref{fig:mass} with the lower two. Any random effect of sampling of the initial conditions may be ruled out because we have done several realisations of each model (hundreds in the case of 1k models). We explain this increment of BHs by the stellar evolution in binary stars. 
The effect of delayed SN explosions of massive stars has been systematically studied, for example by \citet{1992ApJ...391..246P} and \cite{2003NewA....8..817D}, or more recently by \citet{zapartas2017}.
In our models, especially in those compact and not very populous clusters, high-mass binaries are easy to form and difficult to destroy.
We demonstrate this in Fig.~\ref{fig:bhbin} where we plot the mean lifetime of a BH progenitor in a binary before becoming a BH. In the most compact clusters (top panel), especially for a low number of stars, the binary evolution is clearly more significant than in more massive or larger clusters (middle and bottom panels). Due to mass transfer in the binary, even an initially less massive star (which would have ended as a NS) can accrete enough mass from its more massive companion to eventually become a BH. Therefore, we end up with an additional BH. This effect is not as profound in less dense star clusters (lower binary production rate) or in more populous star clusters (higher binary disruption rate).

\subsection{Analytic estimate in comparison to N-body models}

We separate the results into several plots (Fig.~\ref{fig:analytic_1pc}, \ref{fig:analytic_05pc}~\&~\ref{fig:analytic_2pc}) depending on the initial virial radius of the model, which defines the value of the Plummer radius in eq. \eqref{eq:vesc} by $\rp = \frac{3\pi}{16}\,\rv$. The upper limit of the predicted $\rtf$ is calculated as if all the BHs were ejected directly from the centre of the cluster, that is,\ $r = 0$. For the lower limit, we assumed that the ejections take place at the tidal radius determined from eq.~\eqref{eq:tidal} with $M_\mathrm{C} = M_\tot$ (other parameters are the same as in the tidal radius of the models in Sect.~\ref{sec:rtf}). If our simple analytic estimate is correct, we expect the retention fraction determined from the models to be within the shaded area between 0 and $\rt$. For a better understanding of the scale,  in the plots, we also include a dotted line that corresponds to the escape radius of $r = 10\,\pc$ (which is roughly the radius of an expanded star cluster) and a dashdotted line equal to the initial virial radius of each cluster, that is,\ $\rv = 0.5,\ 1.0,$ and $2.0\,\pc$.

In all the figures, we see a general agreement of the analytic estimates and the results from the $N$-body simulations. Especially for the retention fraction in the half-mass radius. The best results are achieved for higher initial masses of the star clusters or wider models (larger $\rv$). This is partially expected from the nature of the escape velocity in eq.~\eqref{eq:vesc}, which is proportional to the square root of the mass of the cluster and inversely proportional to the square root of the initial Plummer radius.

Our analytic prediction does not consider more complicated effects, for example,\ dynamical friction, which can slow the BHs and increase their retention in the cluster (as we have shown in Fig.~\ref{fig:dyn}); the dynamics of core collapse (especially for small clusters $<10$k); or BHs bound in binary systems (mainly in the more compact clusters). Those could very easily shift the limits upwards and yield an even better agreement between $\rtf$ and the results of the numerical models. In the case of higher $\disp$, where the analytic predictions are very strictly giving us $\rtf=0$, we see more fluctuations in the value of $\rtf$ taken from our models with $\rv = 0.5$ or $1.0\,\pc$. Those are also the cases that need more investigation from the point of view of the dynamical effects.

Our analytic estimate has no upper limit for the total mass of the cluster. In Figs.~\ref{fig:analytic_1pc}, \ref{fig:analytic_05pc}~\&~\ref{fig:analytic_2pc} we are showing the results up to $5 \times 10^6\,\msun$ where GCs would be. The trend set by our clusters provides a good indication of how the retention fraction in GCs could behave in those plots. Even further out to the right, outside of these plots, is the mass domain of the UCDs which is documented separately in Fig.~\ref{fig:ucd}.
We note that in the latter case, we use the birth radius according to eq.~(7) from \citet[][]{marks_kroupa}, which is a reasonable assumption since the UCDs are expected to significantly expand to the observed present-day radii \citep{2008MNRAS.386..864D,2010MNRAS.403.1054D}. The retention fraction of BHs in young UCDs should therefore be very high, as shown in Fig.~\ref{fig:ucd}, but self-consistent modelling will be needed to make improved estimates of how the retention fraction evolves as the UCDs expand as a result of the stellar-evolution mass loss.

\section{Conclusions}

The BH retention fraction grows with the increasing initial mass of a star cluster. This is shown by direct $N$-body simulations which corroborate our analytic estimates. Therefore, we conclude that it is possible to estimate the BH retention fraction using a simple analytic formula, according to which the number of escaping BHs is the number of BHs in the tail of the Maxwellian distribution above the escape velocity from the cluster; see eq.~\eqref{eq:rtf_analytic}. This implies that UCDs should have retained more than 80\,\% of their BHs for $\disp \leq 190\,\kms$.

The estimate agrees with our $N$-body results, especially for the retention fraction within the half-mass radius and for star clusters with an initial mass greater than $10^4\,\msun$ and an initial virial radius $\rv \geq 1\,\pc$. In the cases where the analytic prediction does not follow the numerical results, the former serves as a good lower estimate for the retention fraction.
When applying the standard velocity dispersion of SN kicks ($\disp \approx 190\,\kms$), only a few BHs remained bound to the modelled star clusters. Those BHs were either in binaries or they received only a very small kick by chance. With such a high $\disp$, the only places where BHs could be retained is in very massive GCs, UCDs, or an environment comparable to a nuclear cluster. Other possibilities for retaining BHs with high $\disp$ are: a) the kick velocity is to be scaled by a fraction of the envelope that falls back onto the BH, for example,\ from eq.~\eqref{eq:fallback} \citep{Belczynski2002}, or b) they might implode, leaving BH remnants without any kick, that is, putting $p_\mathrm{env+,-} = 0$ in eq.~\eqref{eq:momentum}.
Due to those reasons, if the upper limit of the retention fraction of BHs is about 50\,\% in GCs \citep{peuten,baumgardt_sollima}, then the velocity distribution of the kicks cannot be just one Maxwellian distribution with a velocity dispersion of $\disp \gtrsim 190\,\kms$. This suggests that, if $\disp \gtrsim 50\,\kms$, implosions or some bimodal kick velocity distribution could be valid, as implied for neutron stars by \citet{verbunt}.

Another result we find here is that compact clusters provide an environment in which particularly massive binaries form dynamically and then evolve by binary star evolution to enhance the number of SNe (exploding later than the original SN) providing additional delayed BH formation. Instead of the last SN exploding at $\approx 12\,\Myr$, which is typical for the metallicity and mass range of our models, several BHs needed twice this time to get to the SN stage. This is evident especially in the smallest and most compact clusters, that is,\ 1k stars with $\rv=0.5\,\pc$, although the tendency to later SN explosion is visible in all 1k models, and individual cases of more populous models. The stellar evolution in dynamically formed binary stars (the models here have no primordial binaries) affects mostly the evolution of high-mass stars in small and compact clusters as it is virtually impossible to disrupt them in such systems. 
The other effect of living in a binary for most of the star's life is the overproduction of BHs. An initially relatively less-massive star in a binary system can accrete enough mass to explode as a SN and leave an additional BH that cannot be predicted from the IMF. We have detected this in the case of our 1k model with $\rv=0.5\,\pc$. More populous and less dense clusters also produce dynamically formed binaries, but these do not evolve to produce additional late SNe and BHs.

\begin{acknowledgements}
This study was supported by the Charles University grants SVV-260441 and GAUK-186216.
We thank Sverre Aarseth and Douglas Heggie for the discussion about \texttt{nbody6} and Ond{\v r}ej Chrenko for helpful comments.
We also greatly appreciate access to the computing and storage facilities owned by parties and projects contributing to the National Grid Infrastructure MetaCentrum, provided under the programme ``Projects of Large Research, Development, and Innovations Infrastructures'' (CESNET LM2015042).
We are grateful to the referee for useful comments.
\end{acknowledgements}

\bibliographystyle{aa}
\bibliography{BH-retention-fraction}

\appendix
\onecolumn
\section{Additional Tables \& Figures}


\begin{table*}[!h]
  \centering
  \caption{The BH retention fraction in our models. It is evaluated for each set of the initial conditions (i.e.\ $N$, $\disp$ and $\rv$) in the half-mass radius ($\rh$) and the tidal radius ($\rt$); see eq. \eqref{eq:tidal}, after the last BH has formed ($\tbh$).}
  \begin{tabular}{rd{3}d{2}d{2}d{2}d{2}d{2}d{2}}
    \toprule
    $N$ & \mcc{$\rv\ [\pc]\,:$}   & \mccc{0.5}   & \mccc{1.0}  & \mccc{2.0}  \\%
              & \mcc{$\disp\ [\kms]$} & \mcc{$\rtf (\rh)$} & \mcc{$\rtf (\rt)$} & \mcc{$\rtf (\rh)$} & \mcc{$\rtf (\rt)$} & \mcc{$\rtf (\rh)$} & \mcc{$\rtf (\rt)$} \\%
    \midrule     
    1k   & 3   & 0.23 & 0.43 & 0.14 & 0.28 & 0.05 & 0.24 \\%
         & 10  & -    & -    & 0.01 & 0.03 & -    & -    \\%
         & 25  & -    & -    & 0.01 & 0.02 & -    & -    \\%
         & 50  & 0.05 & 0.08 & 0.00 & 0.00 & 0.00 & 0.00 \\%
         & 190 & 0.04 & 0.05 & 0.00 & 0.00 & 0.00 & 0.00 \\%
         &     &      &      &      &      &      &      \\%
        3k       & 3   & 0.40 & 0.61 & 0.26 & 0.62 & 0.16 & 0.48 \\%
         & 10  & -    & -    & 0.02 & 0.16 & -    & -    \\%
         & 25  & -    & -    & 0.11 & 0.11 & -    & -    \\%
         & 50  & 0.07 & 0.07 & 0.06 & 0.06 & 0.00 & 0.00 \\%
         & 190 & 0.16 & 0.17 & 0.04 & 0.04 & 0.00 & 0.00 \\%
         &     &      &      &      &      &      &      \\%
    10k  & 3   & 0.64 & 0.82 & 0.60 & 0.88 & 0.29 & 0.68 \\%
         & 10  & -    & -    & 0.08 & 0.21 & -    & -    \\%
         & 25  & -    & -    & 0.04 & 0.05 & -    & -    \\%
         & 50  & 0.14 & 0.15 & 0.03 & 0.05 & 0.00 & 0.01 \\%
         & 190 & 0.10 & 0.10 & 0.04 & 0.04 & 0.00 & 0.00 \\%
         &     &      &      &      &      &      &      \\%
    25k  & 3   & 0.76 & 0.84 & 0.80 & 0.98 & 0.38 & 0.82 \\%
         & 10  & -    & -    & 0.30 & 0.45 & -    & -    \\%
         & 25  & -    & -    & 0.06 & 0.09 & -    & -    \\%
         & 50  & 0.09 & 0.16 & 0.00 & 0.00 & 0.00 & 0.00 \\%
         & 190 & 0.08 & 0.13 & 0.03 & 0.03 & 0.00 & 0.00 \\%
         &     &      &      &      &      &      &      \\%
    50k  & 3   & 0.89 & 0.92 & 0.83 & 0.96 & 0.46 & 0.96 \\%
         & 10  & -    & -    & 0.44 & 0.73 & -    & -    \\%
         & 25  & -    & -    & 0.04 & 0.11 & -    & -    \\%
         & 50  & 0.03 & 0.03 & 0.03 & 0.05 & 0.00 & 0.00 \\%
         & 190 & 0.04 & 0.04 & 0.00 & 0.00 & 0.00 & 0.00 \\%
         &     &      &      &      &      &      &      \\%
    100k & 3   & 0.87 & 0.96 & 0.80 & 0.99 & 0.63 & 0.97 \\%
         & 10  & -    & -    & 0.56 & 0.87 & -    & -    \\%
         & 25  & -    & -    & 0.09 & 0.22 & -    & -    \\%
         & 50  & 0.03 & 0.06 & 0.02 & 0.05 & 0.00 & 0.00 \\%
         & 190 & 0.02 & 0.03 & 0.01 & 0.01 & 0.00 & 0.01 \\%
    \bottomrule
  \end{tabular}
  \label{tab:rtf}
\end{table*}

\clearpage


\begin{figure*}
  \centering
  \includegraphics{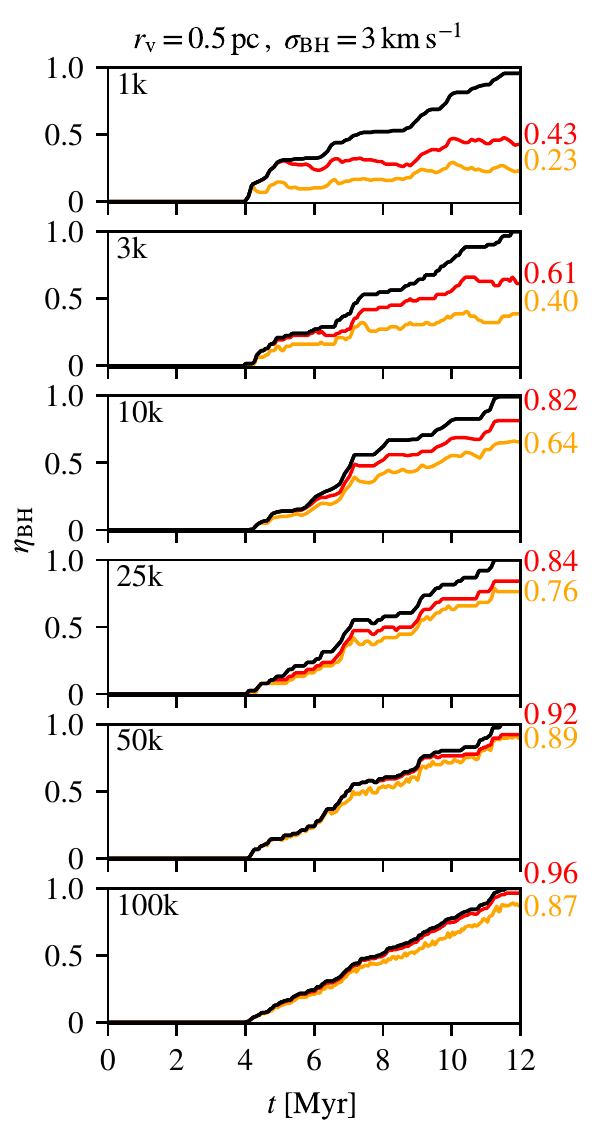}\hfill
  \includegraphics{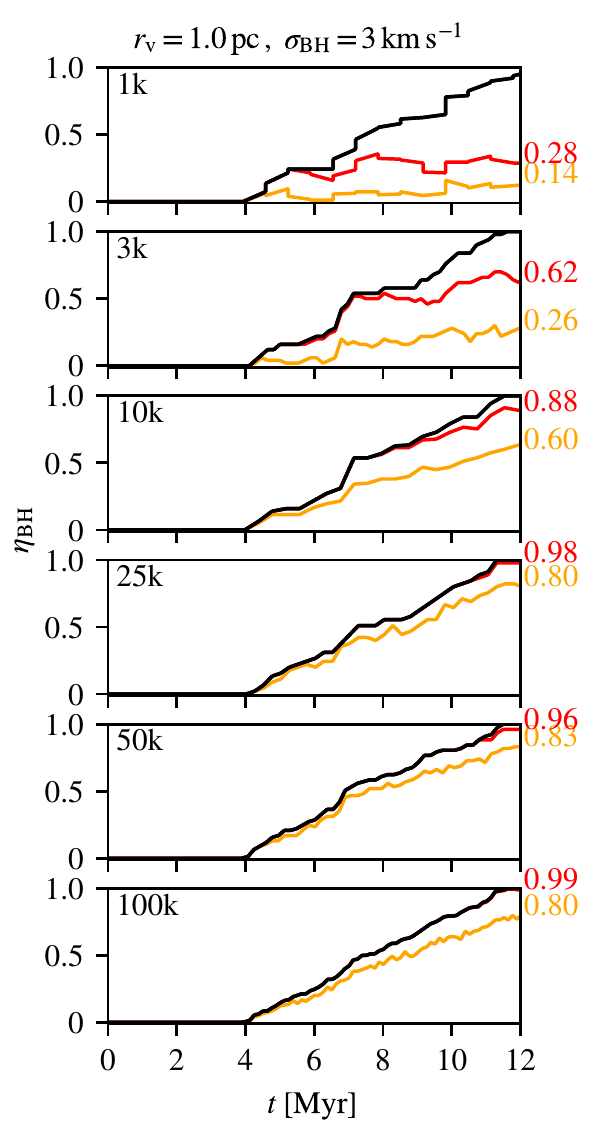}\hfill
  \includegraphics{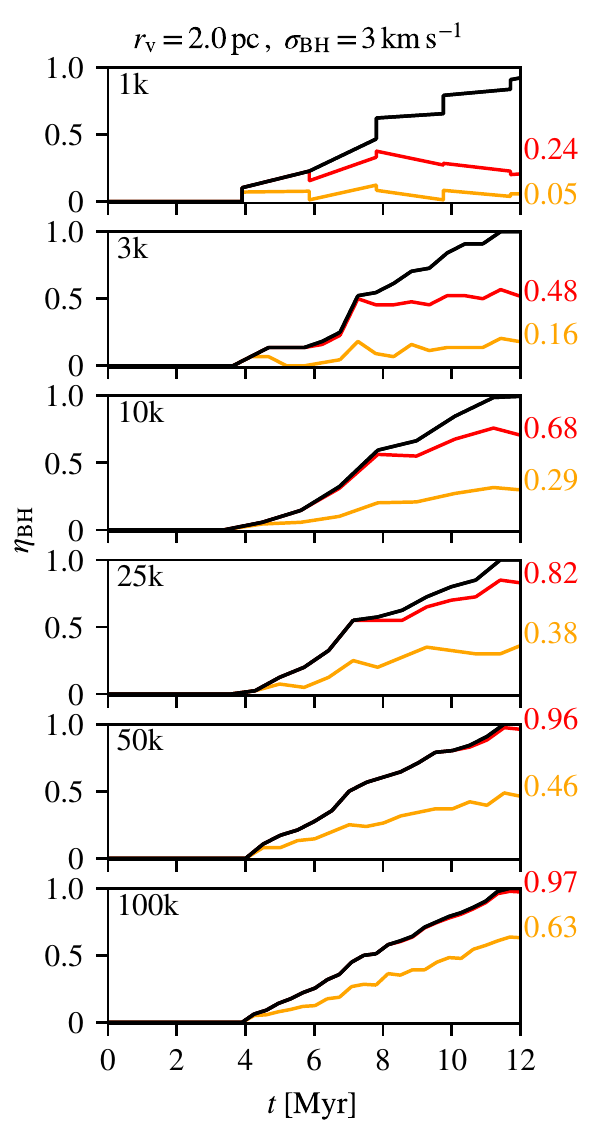}
  
  \includegraphics{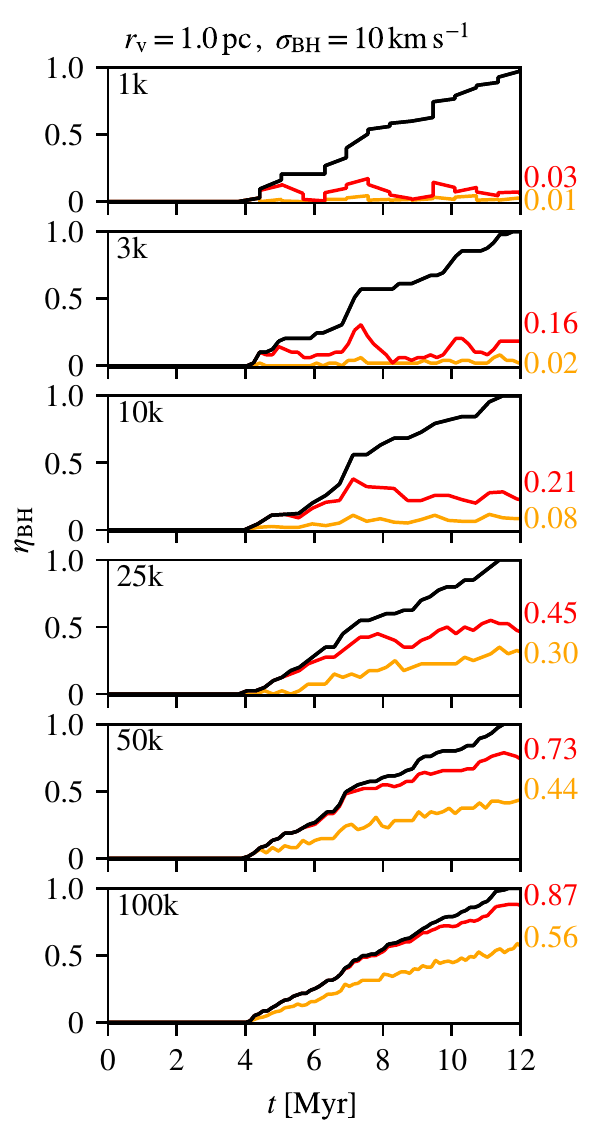}\hfill
  \includegraphics{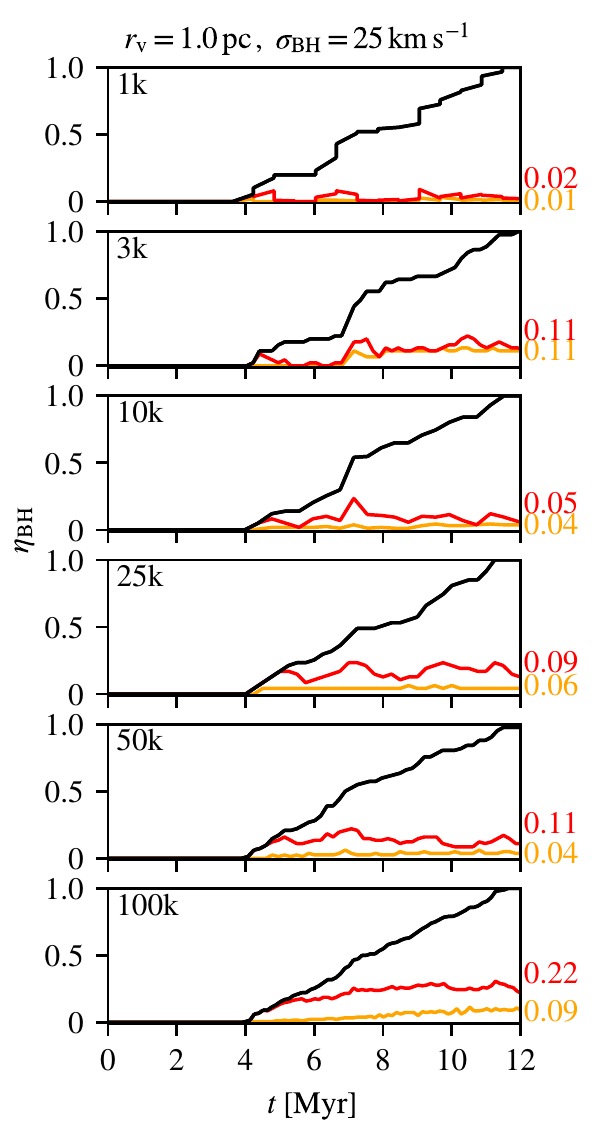}\hfill
  \includegraphics{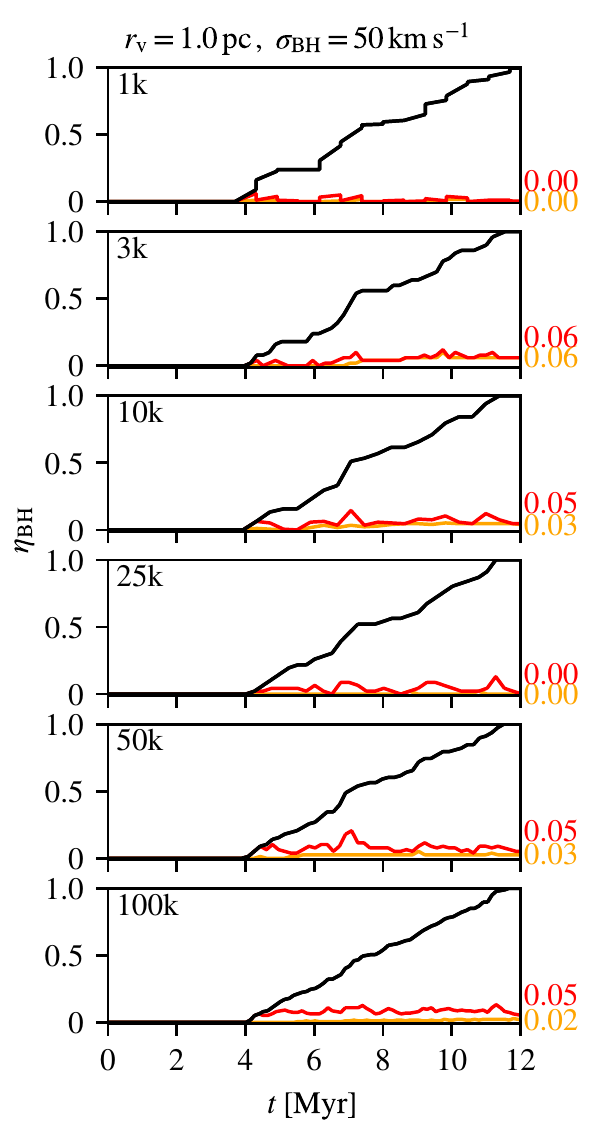}
  \caption{The evolution of the retention fraction, $\rtf$, within the half-mass radius \textbf{(orange)} and the tidal radius \textbf{(red)}. The \textbf{black line} shows the ratio of BHs that have formed with respect to the final number of BHs in the cluster. The curves for 1k and 10k models are averaged over 100 and 10 realisations, respectively.}
  \label{fig:rtf}
\end{figure*}

\begin{figure*}
  \centering
  \includegraphics[width=\linewidth]{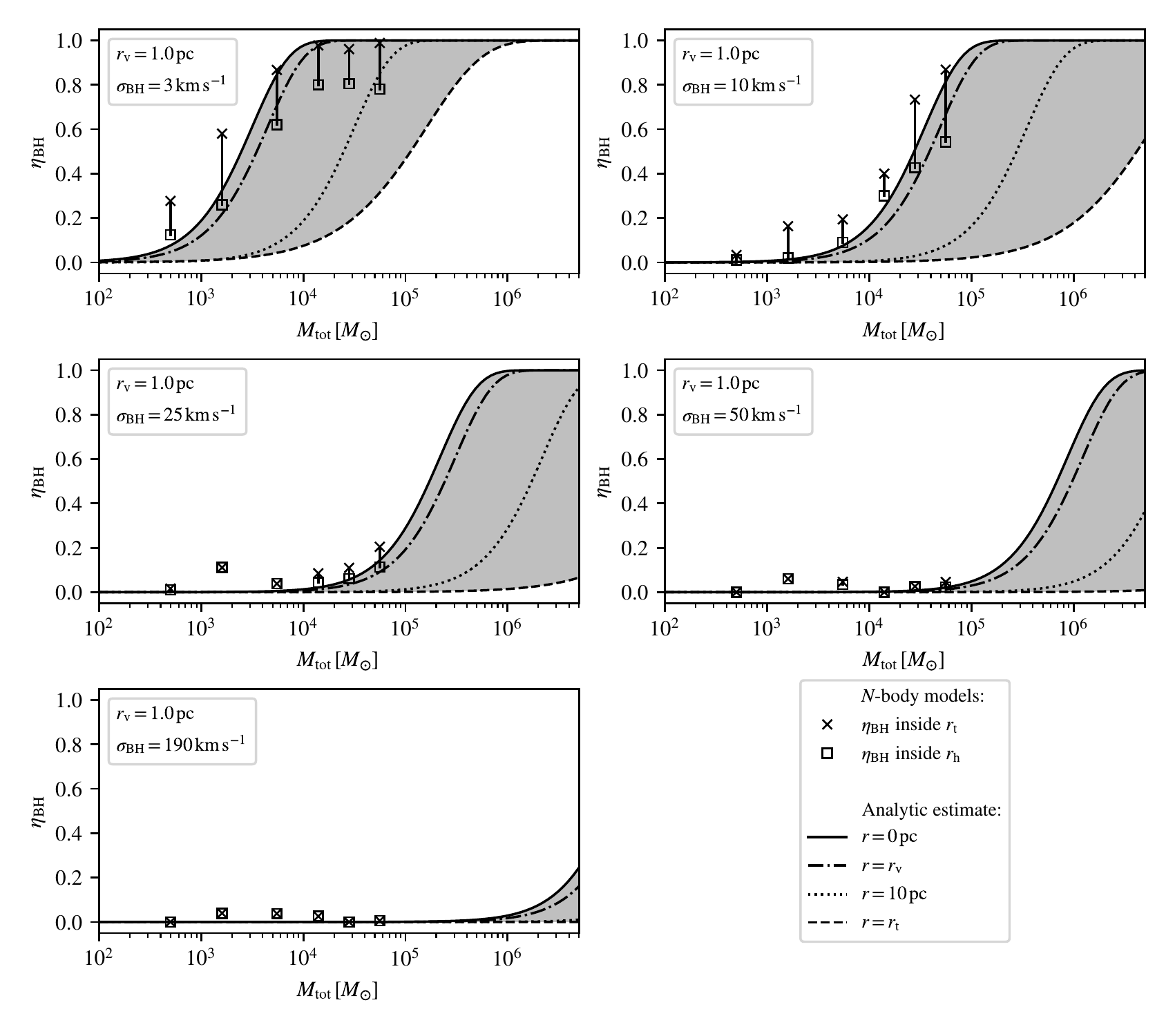}
  
  \caption{A comparison of the retention fraction from the $N$-body models (with $\rv=1.0\,\pc$) with the analytic estimate. The limits from our analytic estimate, eq.~\eqref{eq:vesc}, are represented by a shaded area delimiting the escape radius from 0\,pc (solid line) to the tidal radius (dashed line) determined from eq.~\eqref{eq:tidal}. We also include the curve for an escape radius of 10\,pc (dotted line) and the virial radius (dashdotted line). The squares and crosses represent the retention fraction from our numerical simulations in the half-mass and tidal radius, respectively.}
  \label{fig:analytic_1pc}
\end{figure*}

\begin{figure*}
  \centering
  \includegraphics[width=\linewidth]{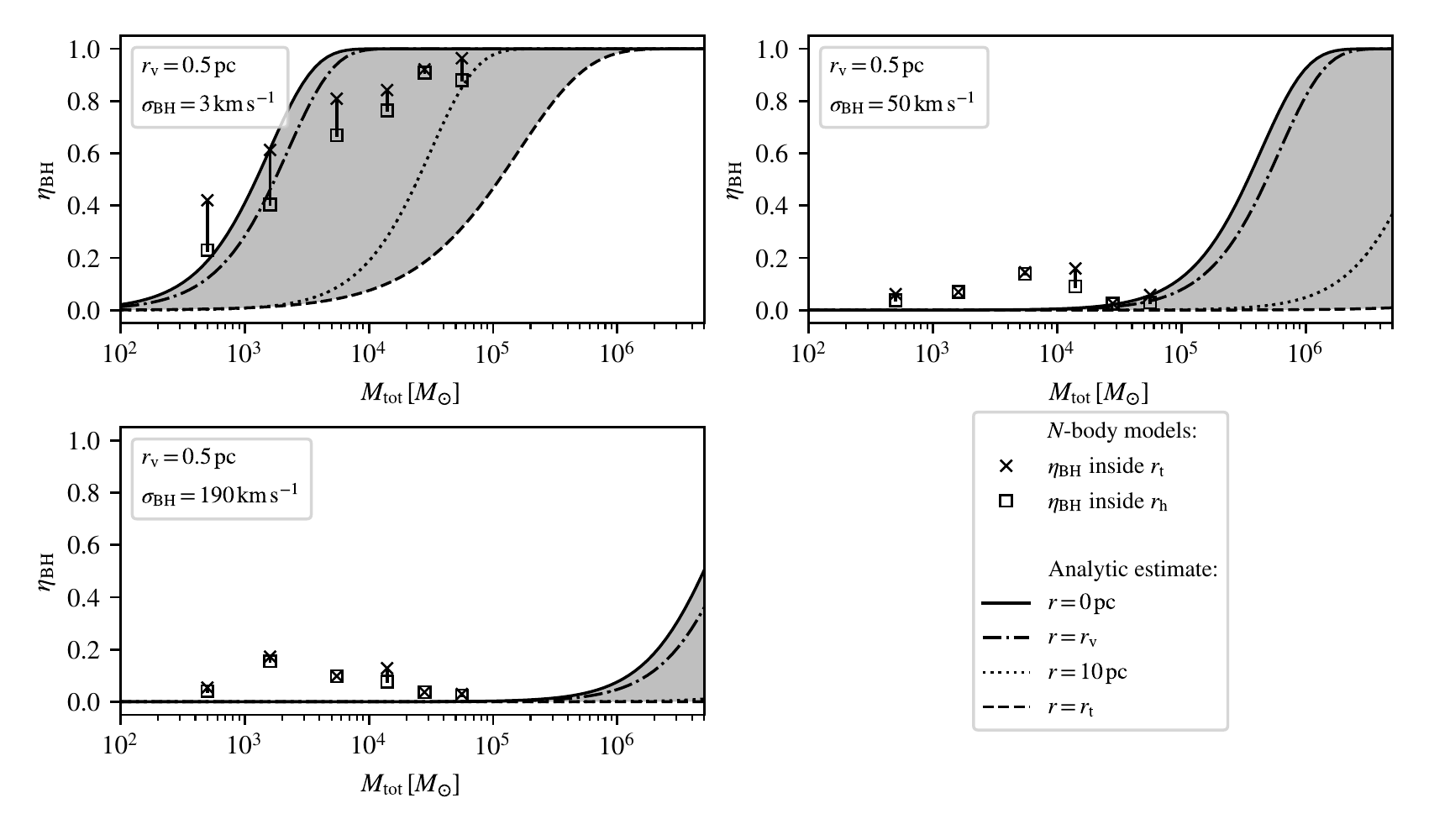}
  
  \caption{As in Fig.~\ref{fig:analytic_1pc} but for the models with the initial virial radius $\rv=0.5\,\pc$.}
  \label{fig:analytic_05pc}
\end{figure*}

\begin{figure*}[p]
  \centering
  \includegraphics[width=\linewidth]{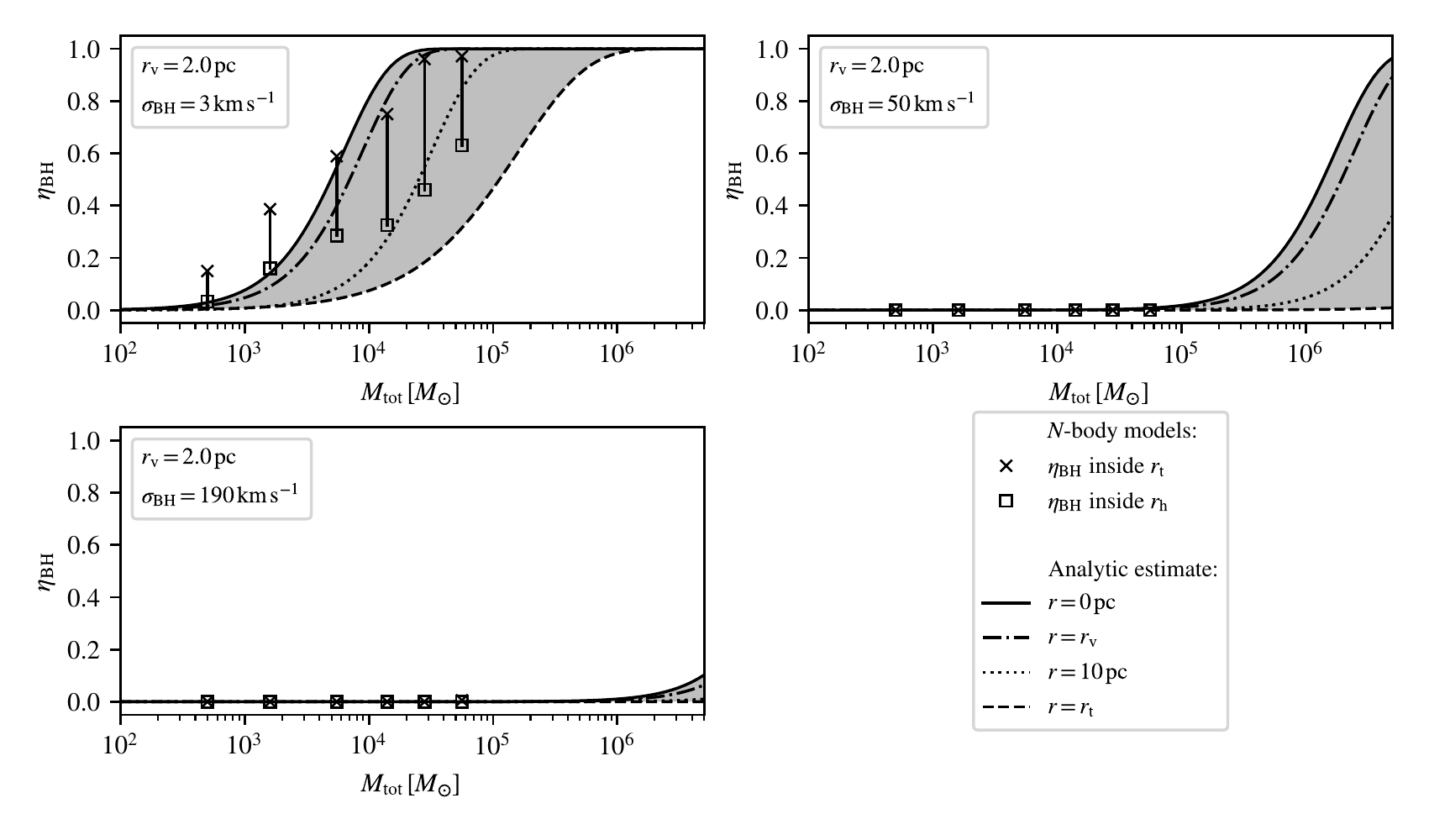}
  
  \caption{As in Fig.~\ref{fig:analytic_1pc} but for the models with the initial virial radius $\rv=2.0\,\pc$.}
  \label{fig:analytic_2pc}
\end{figure*}

\begin{figure*}
  \centering
  \includegraphics[width=\linewidth]{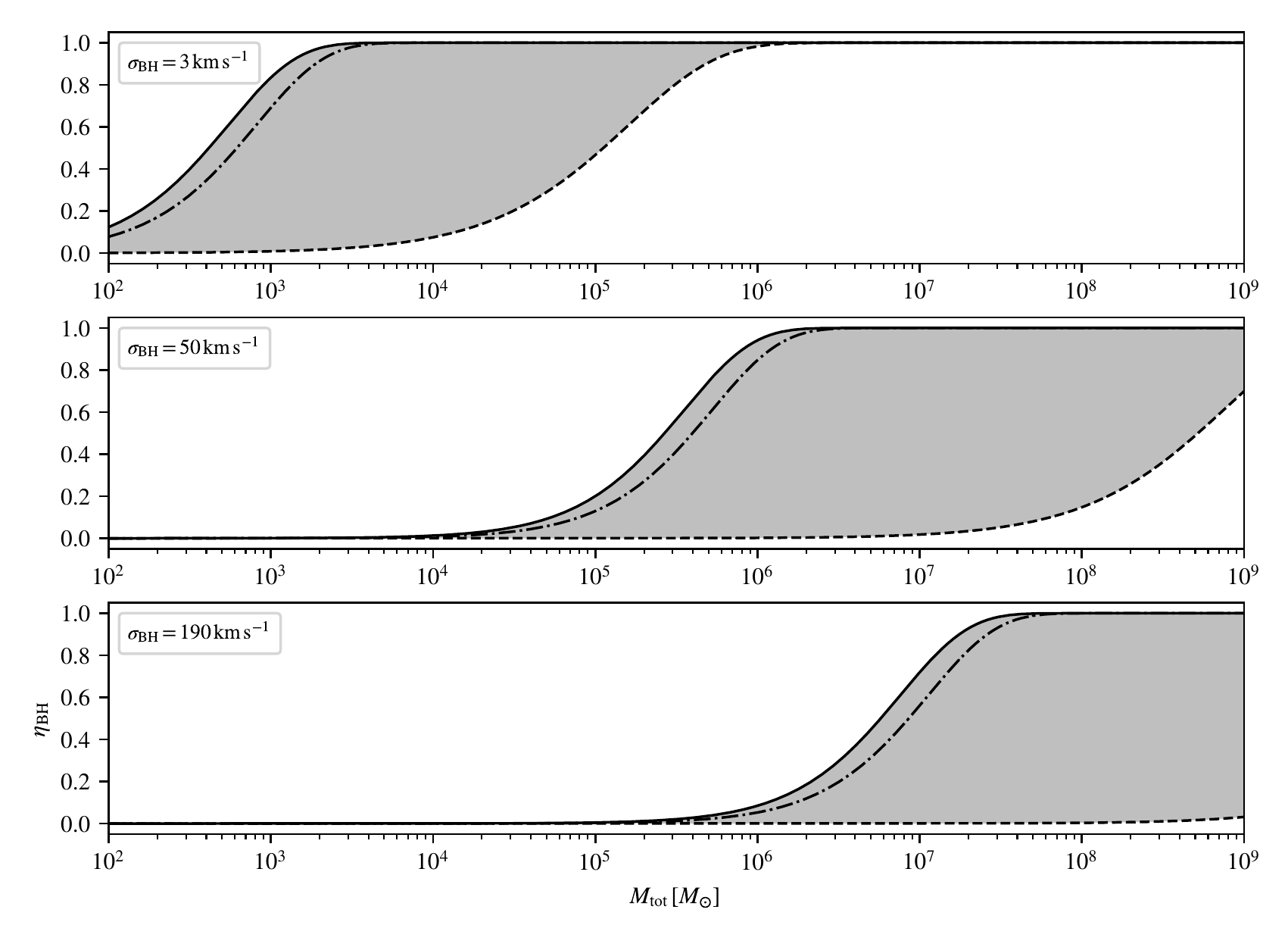}
  
  \caption{The analytic estimate of the retention fraction for a broader range of masses. Here, we use the relation from \citet[][eq.~7]{marks_kroupa} to compute the appropriate radius of the system with a given mass. The escape velocity, eq.~\eqref{eq:vesc}, is calculated from 0\,pc (solid line), the tidal radius (dashed line) determined from eq.~\eqref{eq:tidal}, and the half-mass radius of the model (dashdotted line).}
  \label{fig:ucd}
\end{figure*}

\end{document}